\documentclass{article}
\usepackage{spconf,amsmath,graphicx}

\usepackage{enumitem}
\setlist{nosep, leftmargin=14pt}
\usepackage{bbm}
\usepackage{mwe} % to get dummy images
\usepackage{booktabs}
\usepackage{bm}
\usepackage{multirow}
\usepackage{colortbl}
\usepackage[T1]{fontenc}
\usepackage{array}
\usepackage{tabularray}

\usepackage{amssymb}
\usepackage{adjustbox}
\usepackage{threeparttable,makecell}
\usepackage[table,xcdraw]{xcolor}
\usepackage{xcolor}
\usepackage[colorlinks=true,citecolor=green]{hyperref}
\def \short {} %
\ifx \short \undefined

   \newcommand{\cuthalftablecaptionup}{}
   \newcommand{\cuthalftablecaptiondown}{}
\else

   \newcommand{\cuthalftablecaptionup}{\vspace*{-10pt}}
   \newcommand{\cuthalftablecaptiondown}{\vspace*{-8pt}}
\fi
\title{Unleashing the Infinity Power of Geometry: A Novel Geometry-Aware Transformer (GOAT) for Whole Slide Histopathology Image Analysis}

\name{Mingxin Liu \qquad Yunzan Liu \qquad Pengbo Xu \qquad Jiquan Ma$^{\dagger}$  
\thanks{$^\dagger$ Corresponding author (e-mail: majiquan@hlju.edu.cn)
}}
\address{Department of Computer Science and Technology, Heilongjiang University, China}

\begin{document}
%\ninept
%
\maketitle
\begin{abstract}
The histopathology analysis is of great significance for the diagnosis and prognosis of cancers, however, it has great challenges due to the enormous heterogeneity of gigapixel whole slide images (WSIs) and the intricate representation of pathological features. 
However, recent methods have not adequately exploited geometrical representation in WSIs which is significant in disease diagnosis.
Therefore, we proposed a novel weakly-supervised framework, \textbf{G}e\textbf{o}metry-\textbf{A}ware \textbf{T}ransformer (\textbf{GOAT}), in which we urge the model to pay attention to the geometric characteristics within the tumor microenvironment which often serve as potent indicators. In addition, a context-aware attention mechanism is designed to extract and enhance the morphological features within WSIs. 
% Finally, an effective transformer-enabled framework is introduced for WSI classification which can model long-range dependency. 
Extensive experimental results demonstrated that the proposed method is capable of consistently reaching superior classification outcomes for gigapixel whole slide images.
\end{abstract}


\section{Introduction}
\label{sec:intro}
Pathological slide examination is widely considered the "gold standard" for cancer diagnosis and prognosis, making histopathology whole slide image analysis a crucial deep learning application in clinical practice~\cite{kurc2020segmentation}. In recent years, 
% digital pathology has arisen as a formidable technological advancement in the realm of digitizing whole slide images for the purposes of assessment, storage, and analysis~\cite{dimitriou2019deep,zhang2022gigapixel}. 
the rapid advancement of computational pathology presents an excellent prospect for researchers to design and develop deep learning algorithms for diverse spheres of application, encompassing but not limited to the Gleason grading, survival prediction, and cancer subtyping
~\cite{wang2020weakly,chen2022pan,lu2021data,li2019patch,li2021dual,campanella2019clinical,zhang2022gigapixel,pinckaers2020streaming,shao2021transmil,liu2023mgct}.

The advent of deep learning techniques in recent years has facilitated AI-assisted diagnosis in computational pathology, demonstrating comparable performance to experienced pathologists. Consequently, numerous whole slide image classification techniques based on deep learning have emerged ~\cite{lu2021data,li2019patch,li2021dual,campanella2019clinical,zhang2022gigapixel,pinckaers2020streaming,shao2021transmil}. For example, Ilse et al.~\cite{ilse2018attention} proposed an attention based aggregation operator, giving each instance additional contribution information through trainable attention weights. Li et al.~\cite{li2021dual} designed a novel framework that jointly trained a patch and an image classifier, where the patches are selected softly with instance-level attention. In addition, Shao et al.~\cite{shao2021transmil} presented a transformer-enabled multi-instance learning (MIL) framework to explore both morphological and spatial information among instances. However, these approaches are mostly based on traditional MIL framework and utilize convolutional neural networks (CNNs) as the feature extractor to solely provide limited histology features, which leads to such techniques are not sufficiently context-aware and unable to capture the key morphological feature interactions between cell and tissue that are predictive of cancer diagnosis~\cite{tie2022contextual}.

\begin{figure*}[h]
\centerline{\includegraphics[width=1\textwidth]{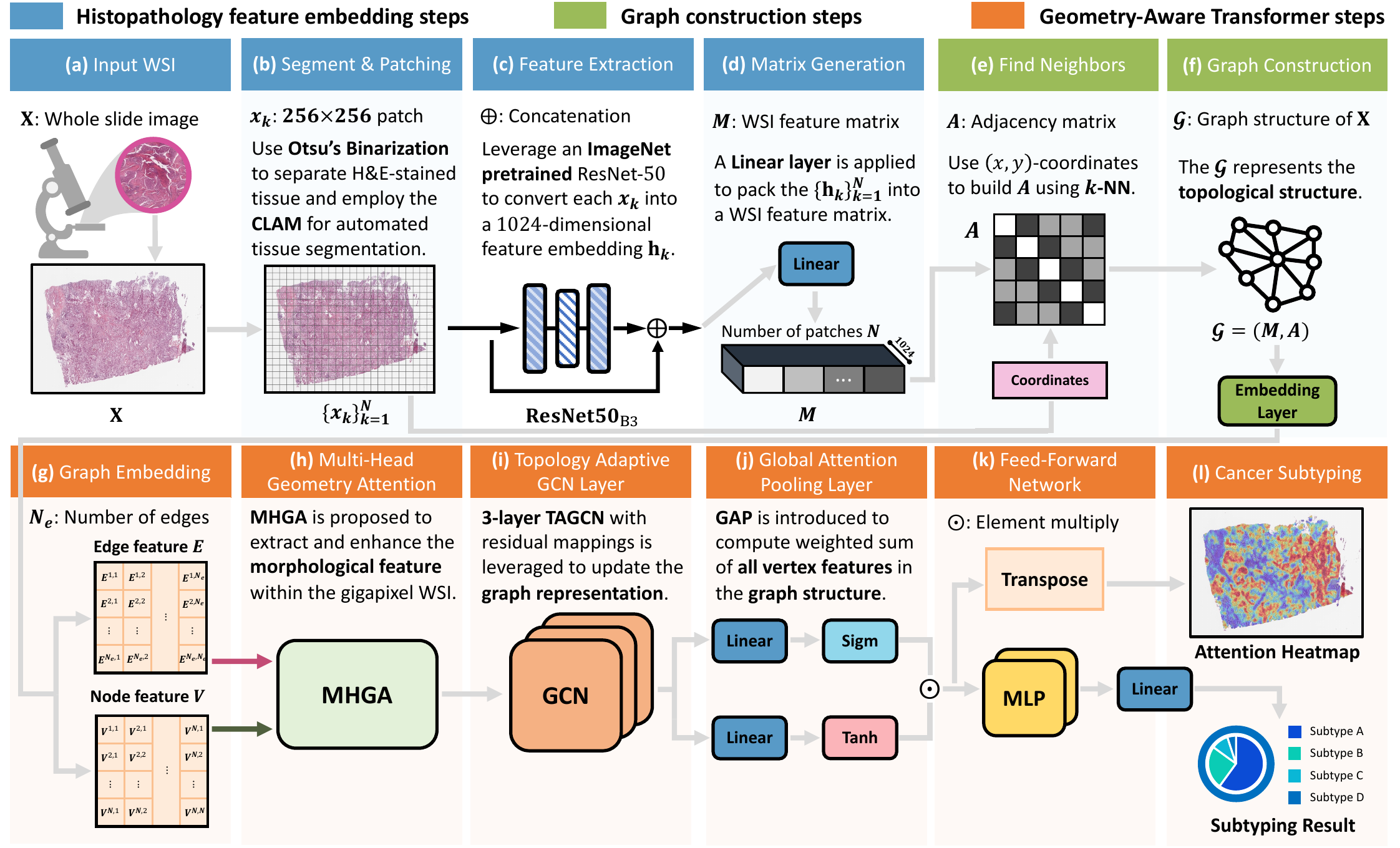}}
\caption{Overview of the proposed \textbf{G}e\textbf{o}metry-\textbf{A}ware \textbf{T}ransformer framework. }
\label{Fig:goat}
\end{figure*}

In this paper, we propose a novel geometry-aware, weakly-supervised, transformer-enabled framework, termed \textbf{G}e\textbf{o}metry-\textbf{A}ware \textbf{T}ransformer (\textbf{GOAT}) as shown in Fig.~\ref{Fig:goat}, to learn contextual histopathological features and complex spatial relationships in the tumor microenvironment for improving the performance of gigapixel WSI classification. The three main contributions of this paper are as follows: (1) We formulated the gigapixel WSI as a fully connected dense graph structure where \textbf{each node represents a non-overlapping patch in the image and each edge represents the spatial relationship between two patches}. (2) We proposed a context-aware attention module (MHGA) to leverage the graph structure to learn the spatial relationships between cell identities and identify the different tissue types. (3) We performed extensive experiments on two benchmarks from TCGA to demonstrate the 
advantages of GOAT over cutting-edge approaches.
% superior performance of our GOAT.
\section{Methodology}

% \subsection{Problem Formulation}
% In a weakly-supervised MIL framework, a dataset that has $N$ bags is typically formulated as $\mathcal{D}=\{ (X_{i}, Y_{i}) \}_{i=1}^{N}$, where $X_{i} = \{ x_{k} \}_{k=1}^{N_{i}}$ denotes the $i$-th bag (WSI) which has $N_{i}$ instances (patches) and $Y_{i}$ is the bag label. The bag label $Y_{i}$ would be positive if and only if there is at least one positive instance in this bag, and it can be defined as follows:
% \begin{equation}
%     Y_{i} =
%     \begin{cases} 
%     1,  & \exists i, y_{i} = 1, \\
%     0,  & \forall i, y_{i} = 0,
%     \end{cases}
% \end{equation}
% where $y_{i}$ is the ground truth label of the $i$-th instance. Therefore, the general \emph{spatial-agnostic} MIL classification process $\mathcal{F}(\cdot)$ can predict the bag label $\hat{Y_{i}}$ without the detailed knowledge and spatial relationship between instances, which can be expressed as follows:
% \begin{equation}
%     \hat{Y_{i}} = \mathcal{F}(X_{i}) = \phi( \eta(f(x_{1}), f(x_{2}), \cdots, f(x_{N_{i}}))),
% \end{equation}
% where $f(\cdot)$ is an instance-level encoder that processes features for each instance independently, $\eta(\cdot)$ is a \emph{permutation-invariant} instance aggregator that aggregates and pools the extracted features to a bag-level feature embedding, and $\phi(\cdot)$ is a bag-level classifier to make final predictions.

\subsection{Graph Construction of Whole Slide Image}

To construct the graph $\bm{\mathcal{G}}$ for the input WSI $\textbf X$, we first utilize Otsu’s Binarization on the saturation channel to separate H\&E-stained tissue from the background and employ the CLAM ~\cite{lu2021data} for automated tissue segmentation. Following segmentation, we extract 256$\times$256 image patches $\left\{x_{k}\right\}_{k=1}^{N}$ without spatial overlapping at the 20$\times$ equivalent pyramid level from all tissue regions identified. To create feature embeddings for the extracted patches, we utilize an ImageNet-pretrained ResNet-50~\cite{he2016deep} as a CNN encoder to convert each patch into a 1024-dimensional feature embedding $\mathbf{h}_{k} \in \mathbb{R}^{1024}$, truncated after the third residual block and adaptive average pooling layer and is then packed into a node feature matrix $\boldsymbol{M} \in \mathbb{R}^{d \times N}$ for the patches in $\textbf X$. 

We construct a $k$-NN graph $\bm{\mathcal{G}} = (\boldsymbol{M}, \boldsymbol{A})$ based on the Euclidean distance between the node features, where $\boldsymbol{A}$ denotes the adjacency matrix for each $\textbf X$ (defined by the coordinates from the tissue segmentation). Finally, we obtain a geometric representation according to the morphological feature of input WSI, which often serves as a powerful cue for cancer diagnosis in clinical pathology practice.

\subsection{Geometry-Aware Transformer}

\smallskip
\noindent{\bf Graph Embedding}
We follow~\cite{chen2023graph} to utilize the graph embedding layer to transform the graph structure as the input of GOAT. As far as we know, besides the vertices, edges also have rich structural information in many types of graphs. Therefore, for exploiting the relationship among nodes and edges in graph data, we transform the graph into edge feature embedding $\boldsymbol{E}$ and node feature embedding $\boldsymbol{V}$ respectively.

\begin{figure}[t]
    \centering
    \centerline{\includegraphics[width=0.5\textwidth,height=0.23\textwidth]{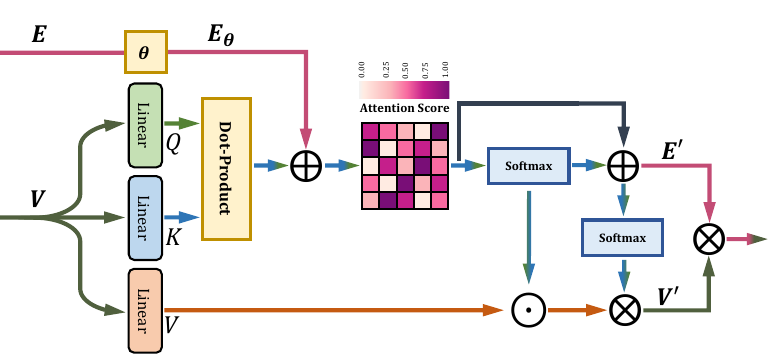}}
    \caption{The proposed Multi-Head Geometry Attention.}
    \label{Fig:mhga}
\end{figure}

\smallskip
\noindent{\bf Multi-Head Geometry Attention}
In recent years, a range of works show global self-attention could serve as a flexible alternative to graph convolution and improve graph representation learning~\cite{ying2021transformers,dwivedi2020generalization}. While most of them only consider part of the graph data and ignore the edge information which is also crucial for geometric structure learning. To this end, we introduce Multi-Head Geometry Attention (MHGA) (shown in Fig.~\ref{Fig:mhga}) to fully model the relationship among vertices and edges and represent the graph structure. We first use linear layer to project the node embedding $\boldsymbol{V}$ to query $\mathrm{Q}$, key $\mathrm{K}$, and value $\mathrm{V}$, 
then we follow~\cite{chen2023graph} to predict layer-specific attention biases $\boldsymbol{E}_{\theta}$ from the edge embedding $\boldsymbol{E}$. After that, we add it to the result of the query-key dot product to generate the attention map $\mathbb{A}$ and update the output node embedding.
This process can be written as follows:
% \begin{equation}
%     \mathcal{A} = \Phi \left(\frac{\mathrm{Q} \cdot \mathrm{K}^{\top}}{\sqrt{d_{head}}} \right) + \boldsymbol{E}_{\theta} = \Phi \left(\frac{\mathrm{Q} \cdot \mathrm{K}^{\top}}{\sqrt{d_{head}}} \right) + \theta \left( \boldsymbol{E} \right) \textbf{W}_{\theta}
% \end{equation}
\begin{equation}
    \mathbb{A} = \frac{\mathrm{Q} \cdot \mathrm{K}^{\top}}{\sqrt{d_{head}}} + \boldsymbol{E}_{\theta} = \frac{\mathrm{Q} \cdot \mathrm{K}^{\top}}{\sqrt{d_{head}}} + \theta \left( \boldsymbol{E} \right) \textbf{W}_{\theta}
\end{equation}
where $d_{head}$ refers to the dimension of each head. Thus, the proposed MHGA can be calculated as:
\begin{equation}
    \begin{aligned}
        & \mathbf{MHGA} \left(\boldsymbol{E}, \boldsymbol{V} \right) = \boldsymbol{E}^{\prime} \otimes \boldsymbol{V}^{\prime}  \\
        & = \Big( \mathbb{A} + \Phi \left( \mathbb{A} \right)\Big) \otimes \Big( \Phi \left(\mathbb{A} \right) \odot \mathrm{V} \otimes \Phi \left( \boldsymbol{E}_{\theta} \right)\Big) \longrightarrow \textbf{h}
    \end{aligned}
\end{equation}
where $\otimes$ and $\odot$ refers kronecker product and element-wise multiplication respectively, $\Phi$ denotes the softmax function, $\textbf h$ denotes the result of MHGA. 

\smallskip
\noindent{\bf Graph Convolution Network}
We further design our graph learning module based on Topology Adaptive Graph Convolutional Network (TAGCN)~\cite{du2017topology} to alleviate the computational complexity without information loss.  The TAGCN layer in MHGA can be formulated as follows:
\begin{equation}\label{out_f}
\mathbf{TAGCN} \left( \textbf h \right) = \sum_{f=1}^{3} \Big( \sum_{n=1}^{N}\textbf{GCN}_{n}^{f} \left( \textbf{h} \right) + \textbf{h}_{f-1} \Big) \longrightarrow \textbf{h}_{\mathrm{GCN}}
\end{equation}
where $\textbf{GCN}\left( \cdot \right)$ is the matrix-vector product, $n=1, 2, \cdots ,N$ and $N$ denotes the number of vertices, $f$ refers the number of GCN layers ($f=3$ here), $\textbf{h}_{f-1}$ denotes the result of previous TAGCN layer. In order to alleviate the problems of over-smoothing and vanishing gradient as the network goes deeper, we add the residual mapping to improve the training. 

\smallskip
\noindent{\bf Global Attention Pooling}
We leverage a global attention-based pooling layer $\mathcal{F}_{\mathrm{GAP}}\left( \cdot \right)$~\cite{ilse2018attention} to adaptively computes a weighted summation of all node features in the graph and pool the node feature embedding $\textbf{h}_{\mathrm{GCN}}$ to a WSI-level embedding:
\begin{equation}
    \begin{aligned}
        & \textbf{h}_{\mathrm{GAP}}  = \mathcal{F}_{\mathrm{GAP}} \left( \sum\limits_{i=1}^N \boldsymbol{\alpha}_{i} \right) \cdot \textbf{h}_{\mathrm{GCN}}  \quad where \\
        & \boldsymbol{\alpha}_{i} = 
        \frac{\mathrm{exp} \left\{ \textbf{W} \Big( \mathrm{tanh} \left( \textbf{V} \cdot \textbf{h}_{i}^{\top}\right) \odot \mathrm{sigm} \left( \textbf{U} \cdot \textbf{h}_{i}^{\top}\right) \Big)\right\}}{\sum\limits_{j=1}^N \mathrm{exp} \left\{ \textbf{W} \Big( \mathrm{tanh} \left( \textbf{V} \cdot \textbf{h}_{j}^{\top}\right) \odot \mathrm{sigm} \left( \textbf{U} \cdot \textbf{h}_{j}^{\top}\right) \Big)\right\}}
    \end{aligned}
\end{equation}
where $\textbf{W}$, $\textbf{V}$, and $\textbf{U}$ are trainable weight matrices, $\boldsymbol{\alpha}_{i}$ is the learnable scalar weight for global attention pooling layer. 

Finally, we utilize a two-layer multi-layer perceptron as the feed-forward network (FFN) in transformer with a fully connected linear layer to downsample and aggregate the final cancer subtyping result, which can be calculated as follows:
\begin{equation}
    \textbf{h}_{\mathrm{Final}} = \mathrm{Linear} \Big( \mathrm{FFN} \left( \textbf{h}_{\mathrm{GAP}} \right)\Big)
\end{equation}
where the $\textbf{h}_{\mathrm{Final}} \in \mathbb{R}^{1024 \times \#\mathrm{CLS}}$ refers the final embedding, $\#\mathrm{CLS}$ denotes the number of classes in current dataset.
\section{Experimental Setup}
\subsection{Datasets and Evaluation Metrics}

\smallskip
\noindent{\bf TCGA-NSCLC}
The TCGA-NSCLC (Non-Small Cell Lung Cancer) dataset includes two sub-types of lung cancer, Lung Adenocarcinoma (LUAD) and Lung Squamous Cell Carcinoma (LUSC). 
The dataset contains a total of 993 diagnostic WSIs, including 507 LUAD slides and 486 LUSC slides. 
% The dataset contains a total of 993 diagnostic WSIs, including 507 LUAD slides from 444 cases and 486 LUSC slides from 452 cases. 
% We benchmarked the performance of our model on this dataset for the lung cancer sub-type classification task.

\smallskip
\noindent{\bf TCGA-RCC}
The TCGA-RCC (Renal Cell Carcinoma) dataset includes three sub-types of kidney cancer, Kidney Chromophobe Renal Cell Carcinoma (KICH), Kidney Renal Clear Cell Carcinoma (KIRC), and Kidney Renal Papillary Cell Carcinoma (KIRP). 
The dataset contains a total of 884 diagnostic WSIs, including 111 CRCC slides, 489 CCRCC slides, and 284 PRCC slides. 
% The dataset contains a total of 884 diagnostic WSIs, including 111 CRCC slides from 99 cases, 489 CCRCC slides from 483 cases, and 284 PRCC slides from 264 cases. 
% We benchmarked the performance of our model on this dataset for the classification of these three different kidney cancer types. 

\smallskip
\noindent{\bf Evaluation Metrics}
To evaluate GOAT, we conducted experiments using 10-fold Monte Carlo cross-validation which each dataset was divided into 60/15/25 partitions for training, validation, and testing. The overall accuracy and the area under the curve (AUC) across the testing splits were employed to measure the diagnosis ability of the models. 

\subsection{Implementation Details}
GOAT is implemented using PyTorch 1.13.1 and PyTorch-Geometric (PyG) 2.0.3, the model is trained on a workstation equipped with an NVIDIA Quadro GV100 GPU. We utilize the cross-entropy loss function, and the model parameters are optimized via stochastic gradient descent using the Adam optimizer with a learning rate of 2e-4 and weight decay of 1e-5. 
% We validate the model after every maximum of 200 epochs and use early stopping on the model when the validation loss does not decrease for 20 consecutive validation epochs. 
Our related code will be publicly made available at \url{https://github.com/lmxmercy/GOAT}.

\section{Results and Analysis}

\subsection{Comparison with State-of-the-Art Methods}
The two used datasets, TCGA-NSCLC and TCGA-RCC, share a common characteristic that the large areas of tumor region in the positive slide (average area \textgreater 80\% per slide), making most of the methods have similar performance, and even the most naive baselines max-pooling and mean-pooling also achieved a relatively high performance. However, GOAT outperformed all comparative baselines in two metrics on these two benchmarks by a significant margin. 
% For TCGA-NSCLC, compared with the cutting-edge MIL model, the attention-based multiple instance learning (ABMIL), the performance is improved by 19.41\% and 13.37\% for accuracy and AUC respectively. 
In comparison with the second best-performing baseline, TransMIL~\cite{shao2021transmil}, GOAT achieved a performance increase of 4.32\% and 2.19\% on accuracy and AUC respectively.
Furthermore, the highest performance of GOAT on the TCGA-RCC dataset demonstrated that our
model also excels in the multi-class classification problem.
The superior improvements of the proposed method showcases its ability to discriminate the cancer regions in the tumor microenvironment, which is often crucial for clinical cancer diagnosis.

\begin{table}[]
    \centering
    \caption{Quantitative evaluation on TCGA-NSCLC and TCGA-RCC two datasets. The highest results are in \textbf{bold}.}
    \resizebox{0.48\textwidth}{!}{
    \begin{tabular}{lcccc}
    \toprule
    \multirow{2}{*}{Methods} & \multicolumn{2}{c}{TCGA-NSCLC} &
    \multicolumn{2}{c}{TCGA-RCC} \\
    \cmidrule(r){2-3} \cmidrule(l){4-5}
    & {Accuracy $\uparrow$ } & {AUC $\uparrow$ } &  {Accuracy $\uparrow $} & {AUC $\uparrow $} \\
    \midrule
    Max-pooling & 0.7282 & 0.8401 & 0.9054 & 0.9786  \\
    Mean-pooling & 0.8593 & 0.9463 & 0.9378 & 0.9879  \\
    ABMIL~\cite{ilse2018attention} & 0.7719 & 0.8656 & 0.8934 & 0.9702  \\
    PT-MTA~\cite{li2019patch} & 0.7379 & 0.8299 & 0.9059 & 0.9700  \\
    StreamingCNN~\cite{pinckaers2020streaming} & 0.8692 & 0.9260 & 0.8817 & 0.9660 \\
    DSMIL~\cite{li2021dual} & 0.8058 & 0.8925 & 0.9294 & 0.9841 \\
    CLAM-SB~\cite{lu2021data} & 0.8180 & 0.8818 & 0.8816 & 0.9723  \\
    CLAM-MB~\cite{lu2021data} & 0.8422 & 0.9377 & 0.8966 & 0.9799  \\
    TransMIL~\cite{shao2021transmil} & 0.8835 & 0.9603 & 0.9466 & 0.9882  \\
    Zhang et al.~\cite{zhang2022gigapixel} & 0.8785 & 0.9377 & 0.9140 & 0.9740   \\ \midrule
    \textbf{GOAT (Ours)} & \textbf{0.9217} & \textbf{0.9813} & \textbf{0.9705} & \textbf{0.9956} \\
    \bottomrule
    \end{tabular}}
	\cuthalftablecaptionup
	\cuthalftablecaptiondown
	\label{tab:results}
\end{table}

\subsection{Ablation Study}
To validate the effectiveness of our designed modules like MHGA, TAGCN, residual mapping, and GAP, we add these components one by one to the baseline. Experimental results are shown in Table~\ref{ablation}. As can be seen, the proposed MHGA has significant improvement (14.29\% on accuracy on TCGA-NSCLC) for cancer subtyping which demonstrated the necessity of the introduced graph structure. In addition, the introduction of TAGCN is also beneficial for WSI classification, performance dramatically degrades for two evaluation metrics when it was removed. Then we add the residual mappings to Model B, the experiment result shows that it plays an essential role in GOAT. We further add a global attention pooling layer, and 2.56\% and 2.07\% improvement is observed for cancer subtyping on accuracy and AUC, respectively.

% We conducted ablation studies on TCGA-NSCLC to verify the effectiveness of the proposed modules. For this purpose, we created a basic MIL model as the baseline (Model A). 

% \smallskip
% \noindent{\bf Graph Convolution Network}
% To evaluate the effectiveness of graph convolutional network layers in the proposed GOAT, we designed Model B by adding GCN layers to Model A. Model B outperforms Model A for 14.39\% on accuracy and 7.06\% on AUC respectively, demonstrating that the introduction of geometric representation learning is crucial in improving the performance of whole slide image classification.

% \smallskip
% \noindent{\bf Dense Connection \& Residual Mapping}
% We further add the dense connections and residual mappings to Model B to obtain the Model D and Model E, the experiment result in Table~\ref{ablation} shows that these two connections play an essential role in GOAT. When the dense connections and residual mappings are removed (Model B), performance dramatically degrades for the accuracy and AUC two evaluation metrics.

% \smallskip
% \noindent{\bf Gated Attention Pooling}
% We also evaluate the effectiveness of gated attention pooling by adding it to Model D, then we compare the resulting Model E and Model D (w/o GAP). The performance comparison proves that the gated attention pooling is beneficial for performance (2.56\% accuracy improvement). This emphasizes the importance of the gated attention pooling in the overall effectiveness of GOAT.

\newcommand{\myvspace}{1pt}
\newcommand{\myboxsize}{0.48\textwidth}
\newcommand{\myarraystretch}{1.3}

\begin{table}[]
	\centering
	\caption{Ablation study on TCGA-NSCLC. MHGA: Multi-Head Geometry Attention. TAGCN: topology adaptive graph convolution network. Residual: residual connections in GCN layers. GAP: global attention pooling layer.}\label{ablation}
	\vspace{\myvspace}
	\renewcommand{\arraystretch}{\myarraystretch}
	\setlength\tabcolsep{6pt}
	\resizebox{\myboxsize}{!}
	{\begin{tabular}{ccccccc}
    \toprule[1pt]
    \multirow{2}{*}{Model} & \multicolumn{4}{c}{{Designs in GOAT}} & \multicolumn{2}{c}{TCGA-NSCLC}\\
    \cmidrule(lr){2-5} \cmidrule(lr){6-7}
    & MHGA & TAGCN & Residual & GAP & Accuracy $\uparrow$ & AUC $\uparrow$ \\ \midrule
    A &            &            &            &            & 0.7532 & 0.8703 \\
    B & \checkmark &            &            &            & 0.8616 & 0.9317 \\
    C & \checkmark & \checkmark &            &            & 0.8801 & 0.9564 \\
    D & \checkmark & \checkmark & \checkmark &            & 0.8987 & 0.9614 \\ 
    E & \checkmark & \checkmark & \checkmark & \checkmark & \textbf{0.9217} & \textbf{0.9813} \\
	\toprule[1pt]
	\end{tabular}}
\end{table}

\subsection{Interpretability and Attention Visualization}
\begin{figure}[t]
\centerline{\includegraphics[width=0.5\textwidth]{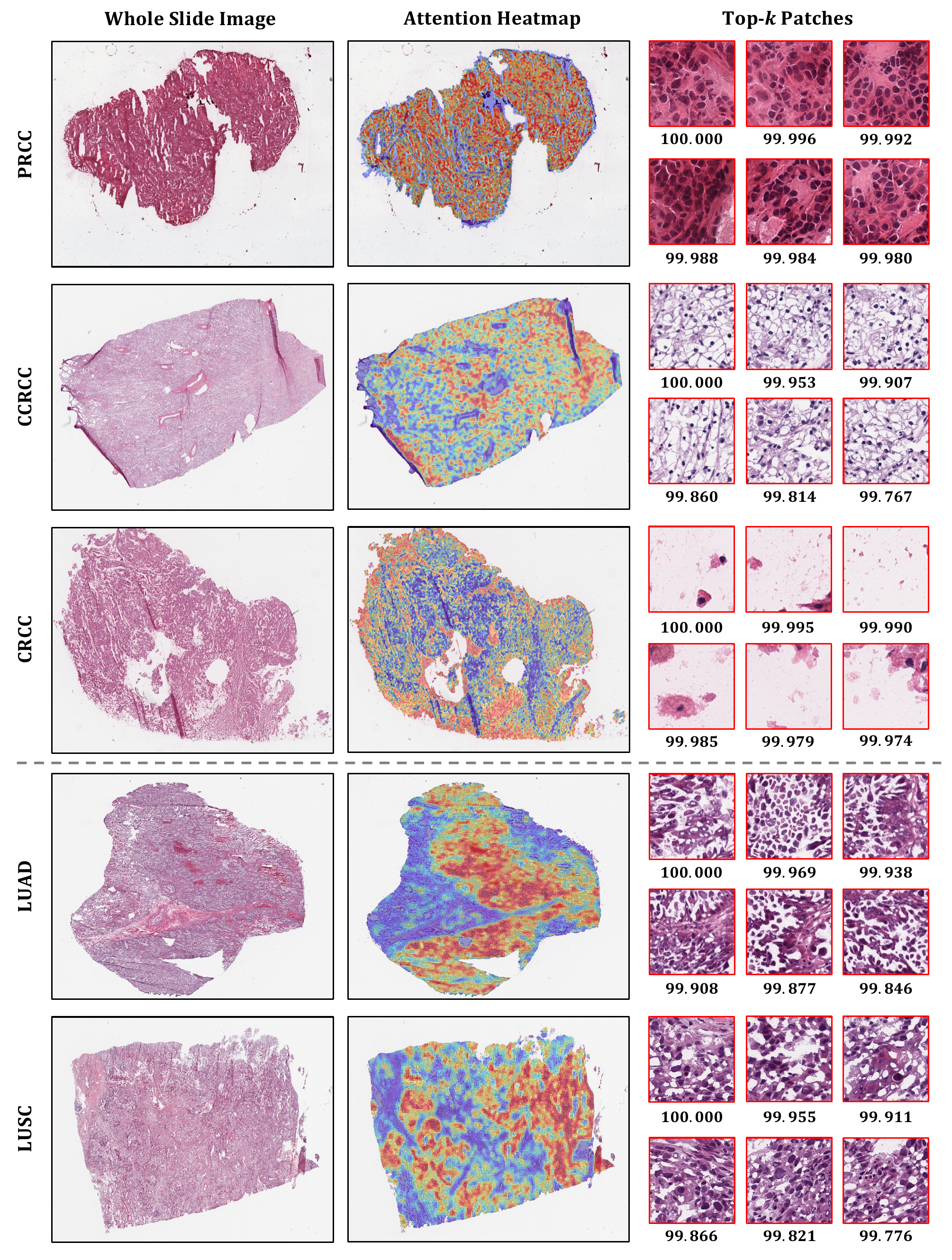}}
\caption{Interpretability and visualization for NSCLC and RCC subtyping. From left to right, the first, second and third column denotes original slide, the attention heatmap, and the top-$k$ highly attended patches with attention scores respectively. 
% The final column shows the top-$k$ patches of the most highly attended regions (red border) and its responding attention scores.
}
\label{Fig:heatmap}
\end{figure}

We demonstrate the interpretability of GOAT by visualizing the attention scores as heatmaps, which indicate the locations of the discriminative regions. 
% We first computed and saved the unnormalized attention scores (before they were converted to probability distribution by applying the softmax function after the linear layer) 
% for all of the $256\times256$ patches extracted from the WSI, then we converted and scaled the attention scores to percentile scores between 0 and 1 (with 1 being most attended and 0 being least attended). Finally, 
We generate an attention heatmap by converting the attention scores for the predicted class of the model into percentiles and mapping the normalized scores to their corresponding spatial location in the original slide. Additionally, we visualize the top-$k$ patches of the most highly attended regions in the input slide which can generally exhibit the corresponding tumor morphology pattern. As shown in Fig.~\ref{Fig:heatmap}, the proposed GOAT can accurately predict the subtyping and localize the related regions in different cancer types which demonstrates that GOAT has the potential to be used for meaningful WSI interpretability and visualization in cancer subtyping problems for clinical practice.

\section{Conclusion}
In this paper, we present an innovative geometry-aware, weakly-supervised, transformer-enabled framework called \textbf{G}e\textbf{o}metry-\textbf{A}ware \textbf{T}ransformer (\textbf{GOAT}) for cancer diagnosis in computational pathology. Our GOAT leverages multi-head geometry attention (MHGA) to exploit the geometric representation from pathological morphological features to improve cancer subtyping performance significantly. Extensive experiments on TCGA-NSCLC and TCGA-RCC two public datasets demonstrate the advantages of the proposed GOAT over cutting-edge methods in improving WSI classification performance. Further ablation study validates the effectiveness of the proposed modules.

\bibliographystyle{IEEEbib}
\bibliography{refs}

\end{document}